\documentclass[journal]{IEEEtran}

\ifCLASSINFOpdf
\else
   \usepackage[dvips]{graphicx}
\fi
\usepackage{url}
\usepackage{amsmath}
\usepackage{epsfig}
\usepackage{amssymb}
\usepackage{svg}
\usepackage{subcaption}
\usepackage{color}
\usepackage{booktabs}
\usepackage{bm}
\usepackage{hyperref}
\usepackage{tabularx}
\usepackage{color}
\hypersetup{hidelinks,
	colorlinks=true,
	allcolors=black,
	pdfstartview=Fit,
	breaklinks=true}

\usepackage{graphicx}

\begin{document}

\title{OMR-NET: A Two-Stage Octave Multi-Scale Residual Network for Screen Content Image Compression}

\author{Shiqi Jiang, Ting Ren, Congrui Fu, Shuai Li \IEEEmembership{Senior Member, IEEE}, Hui Yuan \IEEEmembership{Senior Member, IEEE}
\thanks{This work was supported in part by the National Natural Science Foundation of China under Grants 62222110 and 62172259, the Taishan Scholar Project of Shandong Province (tsqn202103001), and the Natural Science Foundation of Shandong Province under Grant ZR2022ZD38, and the High-end Foreign Experts Recruitment Plan of Chinese Ministry of Science and Technology under Grant G2023150003L. (\textit{Corresponding author: Hui Yuan.})

Shiqi Jiang is with the School of software, Shandong University, Jinan, Shandong, China(e-mail: shiqijiang@mail.sdu.edu.cn).

Congrui Fu is with the School of information science and engineering, Shandong University, Qingdao, Shandong, China(e-mail: fucongrui@mail.sdu.edu.cn).

Ting Ren, Shuai Li and Hui Yuan are with the School of Control Science and Engineering, Shandong University, Jinan, Shandong, China (e-mail: rt630388191@163.com; shuaili@sdu.edu.cn; huiyuan@sdu.edu.cn).}}

\markboth{Journal of \LaTeX\ Class Files, Vol. 14, No. 8, August 2015}
{Shell \MakeLowercase{\textit{et al.}}: Bare Demo of IEEEtran.cls for IEEE Journals}
\maketitle

\begin{abstract}
Screen content (SC) differs from natural scene (NS) with unique characteristics such as noise-free, repetitive patterns, and high contrast. Aiming at addressing the inadequacies of current learned image compression (LIC) methods for SC, we propose an improved two-stage octave convolutional residual blocks (IToRB) for high and low-frequency feature extraction and a cascaded two-stage multi-scale residual blocks (CTMSRB) for improved multi-scale learning and nonlinearity in SC. Additionally, we employ a window-based attention module (WAM) to capture pixel correlations, especially for high contrast regions in the image. We also construct a diverse SC image compression dataset (SDU-SCICD2K) for training, including text, charts, graphics, animation, movie, game and mixture of SC images and NS images. Experimental results show our method, more suited for SC than NS data, outperforms existing LIC methods in rate-distortion performance on SC images. The code is publicly available at \href{https://github.com/SunshineSki/OMR\_Net.git}{https://github.com/SunshineSki/OMR\_Net.git}.
\end{abstract}


\IEEEpeerreviewmaketitle

\section{Introduction}
\label{sec:intro}
\begin{figure}
\begin{minipage}[t]{0.48\linewidth}
  \vspace{0pt}
  \centering
  \centerline{\epsfig{figure=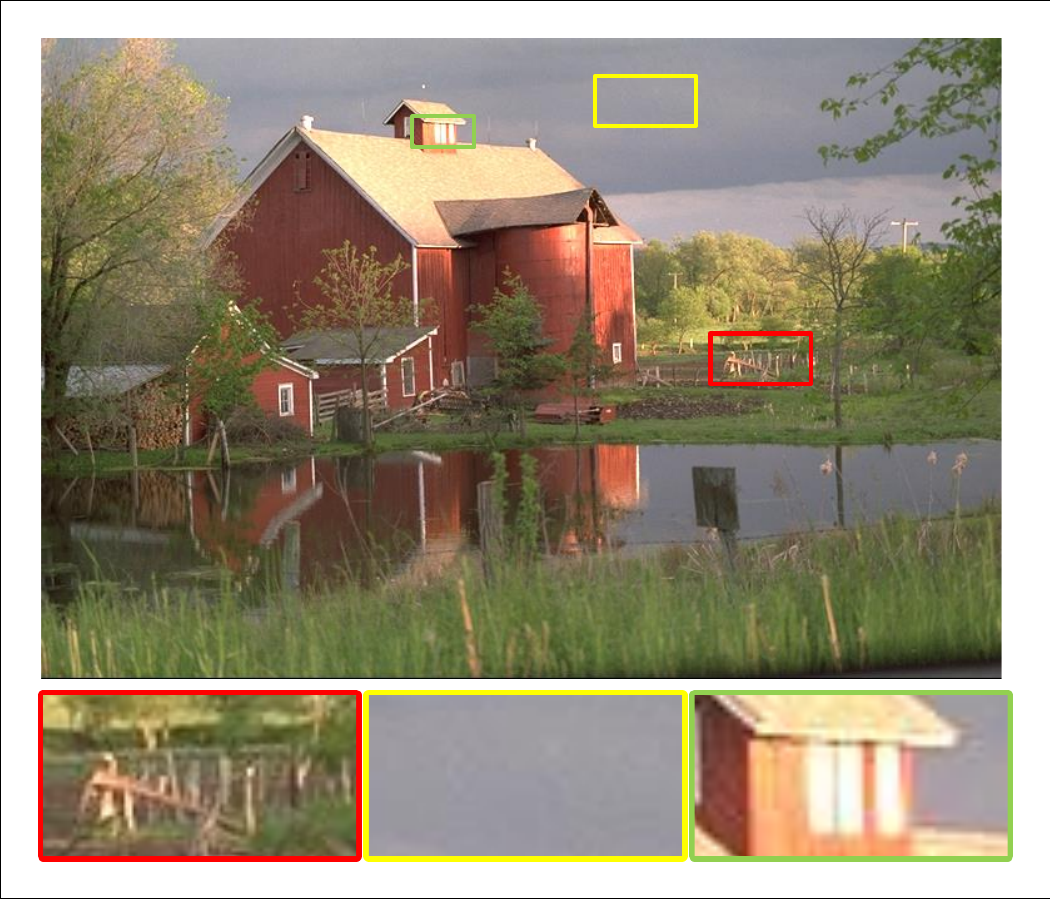,width=4cm}}
  \centerline{(a) NS image}\medskip
\end{minipage}
\begin{minipage}[t]{0.48\linewidth}
  \vspace{0pt}
  \centering
  \centerline{\epsfig{figure=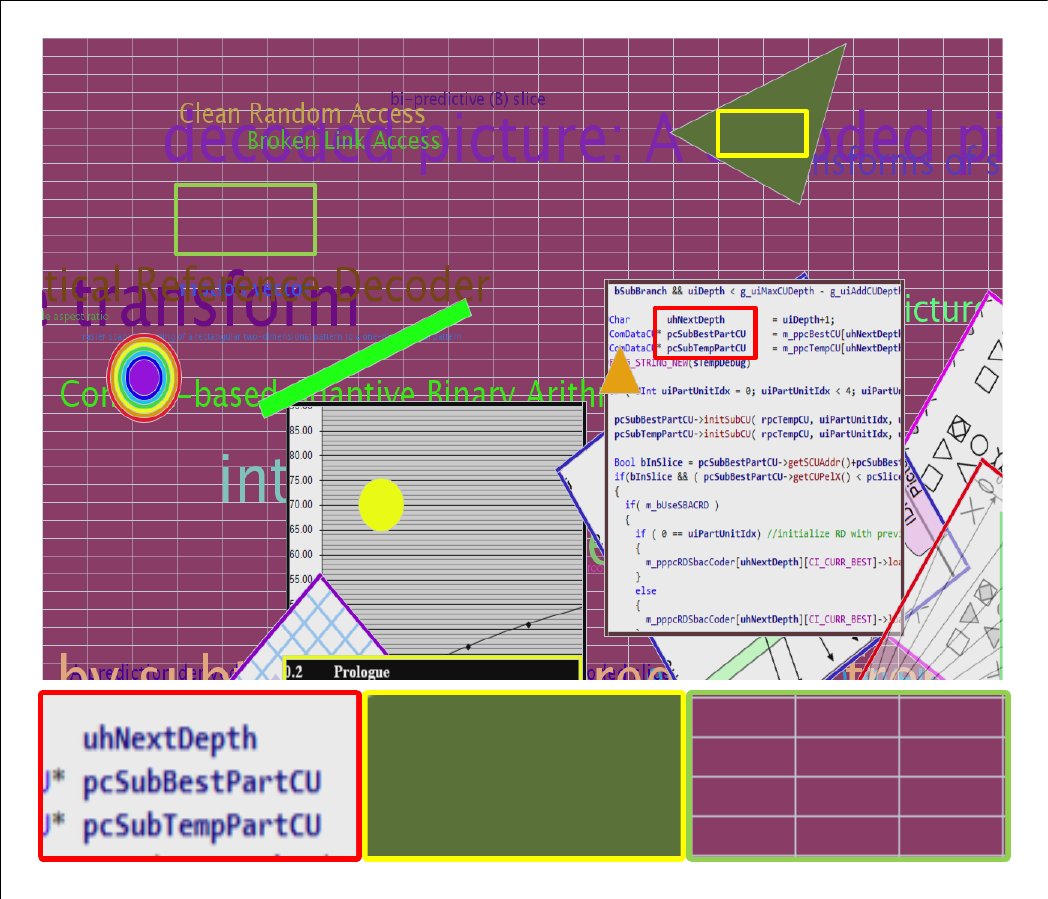,width=4cm}}
  \centerline{(b) SC image}\medskip
\end{minipage}
\caption{Comparison between (a) NS and (b) SC. Red boxes: characteristics between sharp NS textures and sharp SC textures. Yellow boxes: distinctions between smooth NS with visually similar pixels and smooth SC with identical pixels. Green boxes: differences between similar NS patterns and repetitive SC patterns with identical pixels.}
\label{fig:NSI_SCI}
\vspace{-6pt}
\end{figure}
\begin{figure*}
\centering
\includegraphics[width=0.8\linewidth]{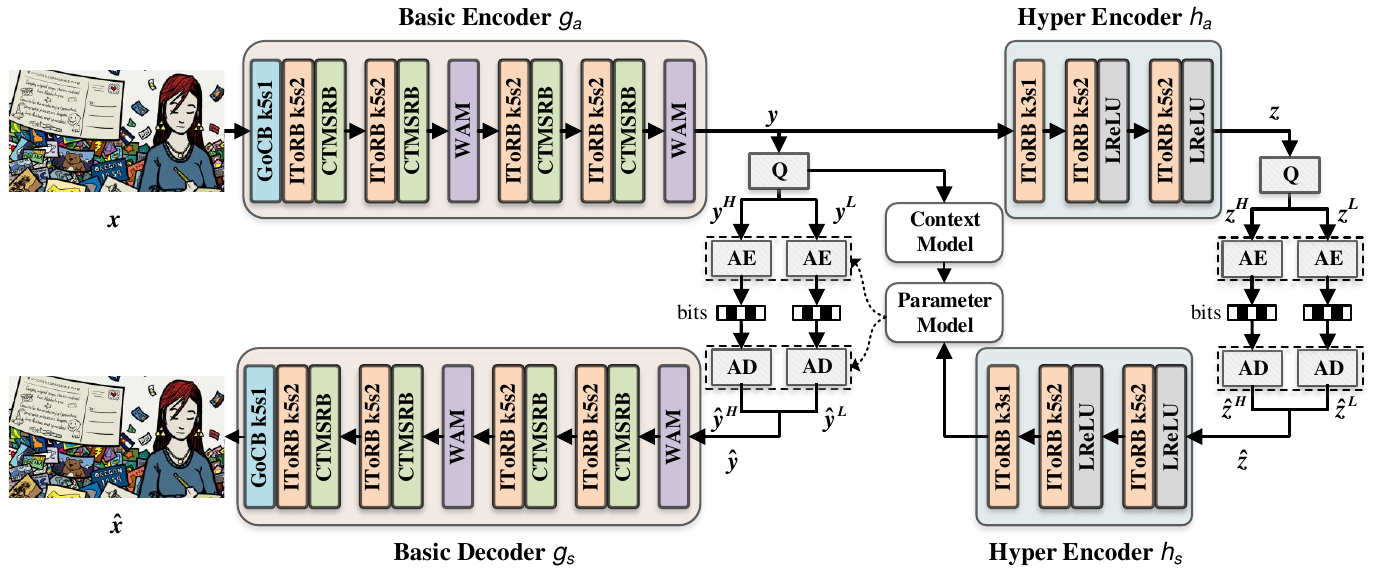}
\caption{Proposed framework. Q denotes quantization, AE and AD denote arithmetic encoding and decoding, respectively. The decoder mirrors the encoder structure, utilizing transposed convolutions.}
\label{fig:framework}
\vspace{-5pt}
\end{figure*}
In the evolving digital era, the rise of internet-based screen content (SC) due to increased online activities, such as remote work and gaming, highlights the need for efficient SC encoding and transmission. Traditional image compression standards like H.265/HEVC and H.266/VVC~\cite{bross2021overview} have been effective, but recent learned image compression (LIC) algorithms already outperform H.266/VVC. However, LIC algorithms are primarily optimized for natural scene (NS) images, limiting their suitability for SC's unique attributes. SC images~\cite{wang2021perceptual,wang2021quality}, which are computer-generated, exhibit distinct properties from NS images, such as high/low-frequency elements, less noise, and repetitive patterns, as shown in Fig. \ref{fig:NSI_SCI}. These differences challenge traditional coding techniques designed for NS images. Many traditional coding standards have integrated screen content coding (SCC), leading to significant advancements, comprehensive solutions for SCC are still lacking, despite the considerable potential of LIC methods~\cite{jiang23,10132493,9044407} in this domain.

In this paper, we propose a LIC method optimized for SC. Our approach involves separate compression of high and low-frequency features using improved octave convolution, capitalizing on its efficacy in feature extraction. Additionally, we propose cascaded two-stage multi-scale residual blocks (CTMSRB) for advanced nonlinear transformation, coupled with a window-based attention module (WAM)~\cite{zou2022devil} to enhance feature extraction in high-contrast regions. Furthermore, we create a dedicated dataset for SC image compression covering various SC types. Experimental results consistently demonstrate the superiority of our algorithm over existing methods, underscoring its efficacy and versatility. 
\section{related work}
\label{sec: related work}
\subsection{Traditional screen content image compression}
Recent video coding standards have introduced SCC tools~\cite{xu2021overview}. Specifically, H.265/HEVC-SCC is an optimized version of H.265/HEVC tailored for screen content coding, incorporating several key coding tools such as intra block copy (IBC), palette mode, adaptive color transform (ACT), and advanced motion vector resolution (AMVR). H.266/VVC further enhances encoding performance by introducing transform skip residual coding (TSRC) and block-based differential pulse-code modulation (BDPCM)~\cite{nguyen2021overview}.
\subsection{Learned screen content image compression}
LIC methods have recently surpassed traditional methods in NS image compression~\cite{10323534,8977394}, while SCC still remains significant research potential~\cite{shen2023dec}. Ballé {\em et al.} first introduced an end-to-end LIC framework \cite{balle2017end} and incorporated a hyperprior structure \cite{balle2018variational}. Wang {\em et al.}~\cite{wang2022transform} innovatively incorporated the concept of transform skip for end-to-end encoding of SC based on \cite{balle2018variational}. Additionally, Heris {\em et al.}~\cite{heris2023multi} boosted rate-distortion (RD) performance by separating SC encoding into synthetic and natural components. Shen {\em et al.}~\cite{shen2023dec} proposed an entropy-efficient transfer learning module in the decoder, aimed at bridging the gap between NS and SC image compression. Tang {\em et al.}~\cite{tang2023feature} employed a super-resolution technique to enhance performance by magnifying decoded SC images. These collective advancements signify the progress in learned SCC techniques, yet they overlook the high and low-frequency differences and repetitive pattern features of SC. Therefore, we propose an LIC framework for SC based on this characteristic.
\vspace{-4pt}
\section{proposed method}
\label{sec: proposed method}
\begin{figure}
\includegraphics[width=\linewidth]{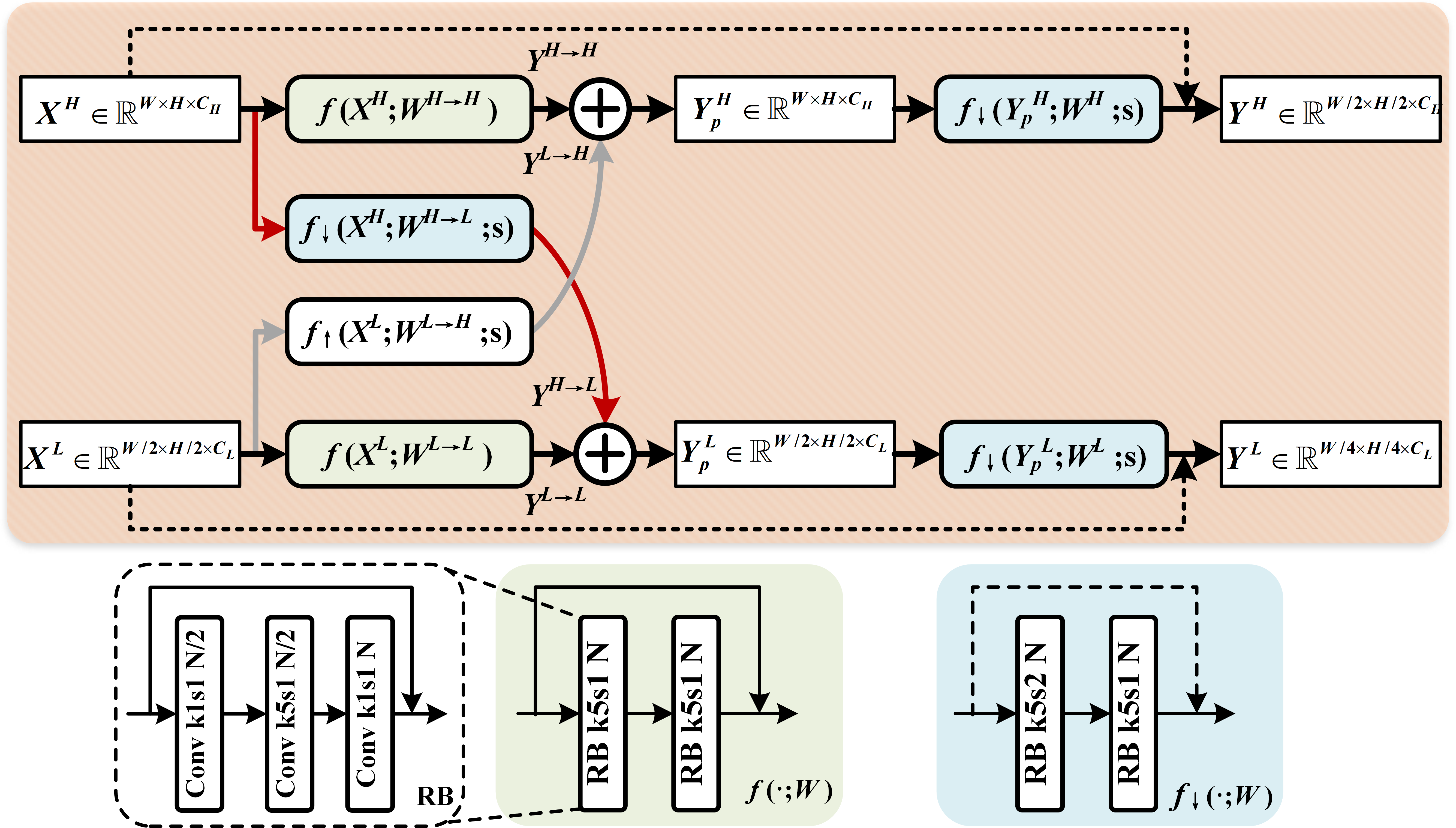}
\caption{The structure of the IToRB. Red arrows indicate high-to-low frequency information transmission, while gray arrows depict the reverse. Dotted arrows represent shortcut with stride. RB denotes residual block, k`m's`n' denotes convolution with a kernel of `m' and a stride of `n'. N denotes the number of channels. The structure of $f$, $f_{\uparrow}$ and $f_{\downarrow}$ are redesigned. $f_{\uparrow}$ is a variant of $f_{\downarrow}$ employing transposed convolution.}
\label{fig:TwoGoOct}
\vspace{-6pt}
\end{figure}
We present an LIC algorithm tailored for SC based on an two-stage octave multi-scale residual network (namely OMR-Net). To our knowledge, this is the first LIC work by exploiting the extreme frequency variation in SC images. The network architecture, as illustrated in Fig. \ref{fig:framework}, is built on octave-based network~\cite{chen2022two} and comprises two fundamental networks: the base network and the hyperprior network. This architecture uses an improved two-stage octave residual block (IToRB), a cascaded two-stage multiscale residual blocks (CTMSRB), and a window-based attention module (WAM) to optimize the compression process. We employ the autoregressive model~\cite{minnen2018joint} to construct the context model for entropy coding, which is implemented using 5×5 masked convolutions.
\vspace{-12pt}
\subsection{Improved two-stage octave residual block}
Akbari {\em et al.}~\cite{akbari2021learned} extended octave convolution (OctConv)~\cite{chen2019drop} as a generalized OctConv (GoConv), which led to redundant downsampling and upsampling operations. Chen {\em et al.}~\cite{chen2022two} introduced a two-stage octave residual block (ToRB) to separate upsampling from downsampling, utilizing OctConv for initial high/low-frequency inputs but using vanilla convolutions inadequate for effective feature extraction. To efficiently capture the high and low-frequency features of SC images, we improve the ToRB and applies it within a SCC framework, as depicted in Fig. \ref{fig:TwoGoOct}. Unlike~\cite{chen2022two}, we initially obtain the high-frequency feature $\boldsymbol{{X}^{H}} \in \mathbb{R}^{{W} \times {H} \times {C}_{H}}$ and low-frequency feature $\boldsymbol{{X}^{L}} \in \mathbb{R}^{{W/2} \times {H/2} \times {C}_{L}}$ via GoConv. Subsequently, we replaced vanilla convolutions with residual blocks in the proposed IToRB to strengthen the depth of the network while avoiding the disappearance of gradients, and therefore enhancing nonlinear representation of the network. The IToRB takes $\boldsymbol{{X}^{H}}$ and $\boldsymbol{{X}^{L}}$ as inputs, and generates corresponding features $\boldsymbol{{Y}^{H}} \in \mathbb{R}^{{W/2} \times {H/2} \times {C}_{H}}$ and $\boldsymbol{{Y}^{L}} \in \mathbb{R}^{{W/4} \times {H/4} \times {C}_{L}}$ with halved resolutions. Here, ${C}_{H} = (1 - \alpha) C$ and ${C}_{L} = \alpha C$, where $\alpha$ represents the ratio of channel allocation to input:
\begin{align}
\boldsymbol{Y^H} &= f_{\downarrow}{(\boldsymbol{Y_p^H}; \boldsymbol{W^H}; s)}+f_{st}{(\boldsymbol{X^H})},
\end{align}
\begin{align}
\boldsymbol{Y^L} &= f_{\downarrow}{(\boldsymbol{Y_p^L}; \boldsymbol{W^L}; s)}+f_{st}{(\boldsymbol{X^L})},\\
\boldsymbol{Y_p^H} &= \boldsymbol{Y^{H \rightarrow H}}+\boldsymbol{Y^{L \rightarrow H}},\\
&=f(\boldsymbol{X^H};\boldsymbol{W^{H \rightarrow H}})+f_{\uparrow}(\boldsymbol{X^L};\boldsymbol{W^{L \rightarrow H}};s),\\
\boldsymbol{Y_p^L} &= \boldsymbol{Y^{L \rightarrow L}}+\boldsymbol{Y^{H \rightarrow L}},\\
&=f(\boldsymbol{X^L};\boldsymbol{W^{L \rightarrow L}})+f_{\downarrow}(\boldsymbol{X^H};\boldsymbol{W^{H \rightarrow L}};s),
\end{align}
where, $\boldsymbol{Y_p^H}$ and $\boldsymbol{Y_p^L}$ represent the output of OctConv, $f(\cdot;\boldsymbol{W})$ represents a residual network with parameter $\boldsymbol{W}$, while $\uparrow$ and $\downarrow$ correspond to up and down sampling with a stride of $s$. $f_{st}(\cdot)$ denotes a skip with stride convolution.
\subsection{Cascaded two-stage multi-scale residual blocks}
\begin{figure}[ht]
\begin{subfigure}{0.22\textwidth}
  \centering
  \includegraphics[width=0.8\linewidth]{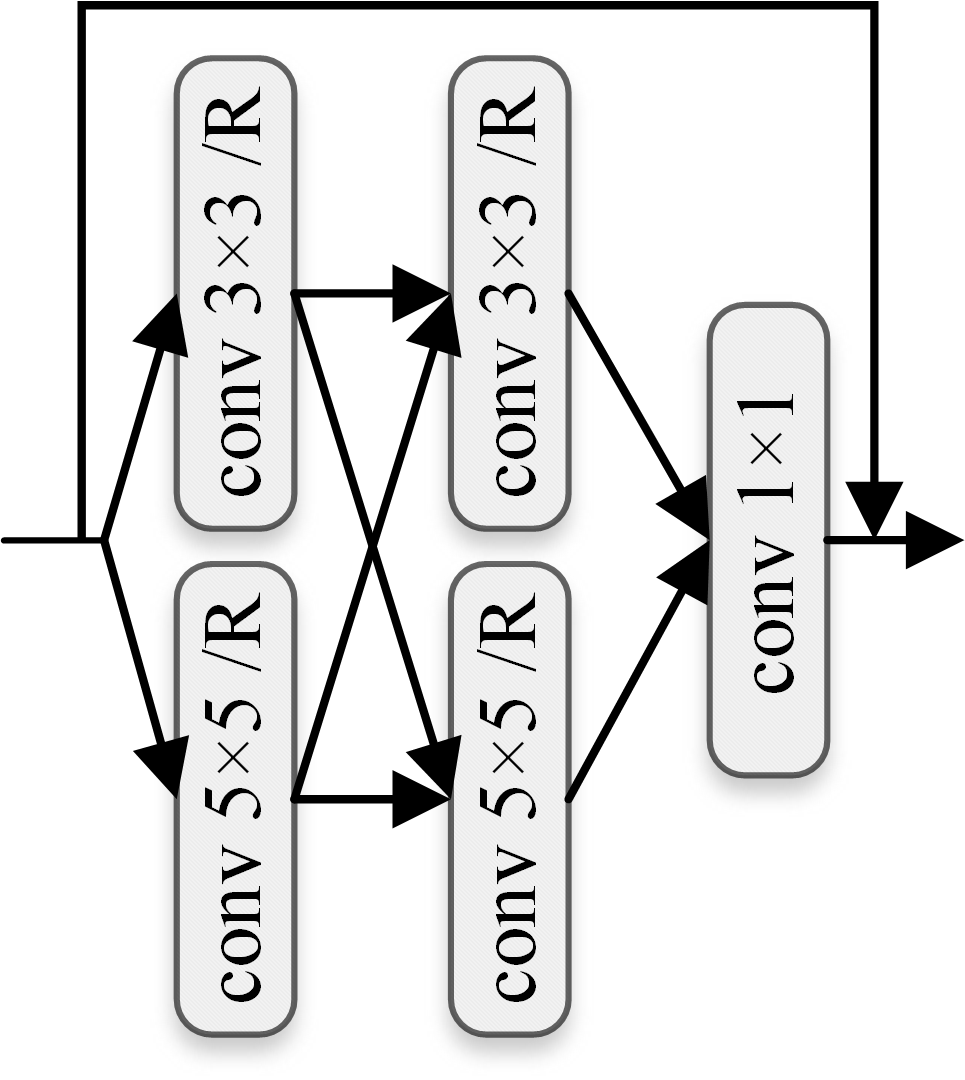}
  \subcaption{MSRB}
\end{subfigure}
\begin{subfigure}{0.26\textwidth}
  \centering
  \includegraphics[width=0.8\linewidth]{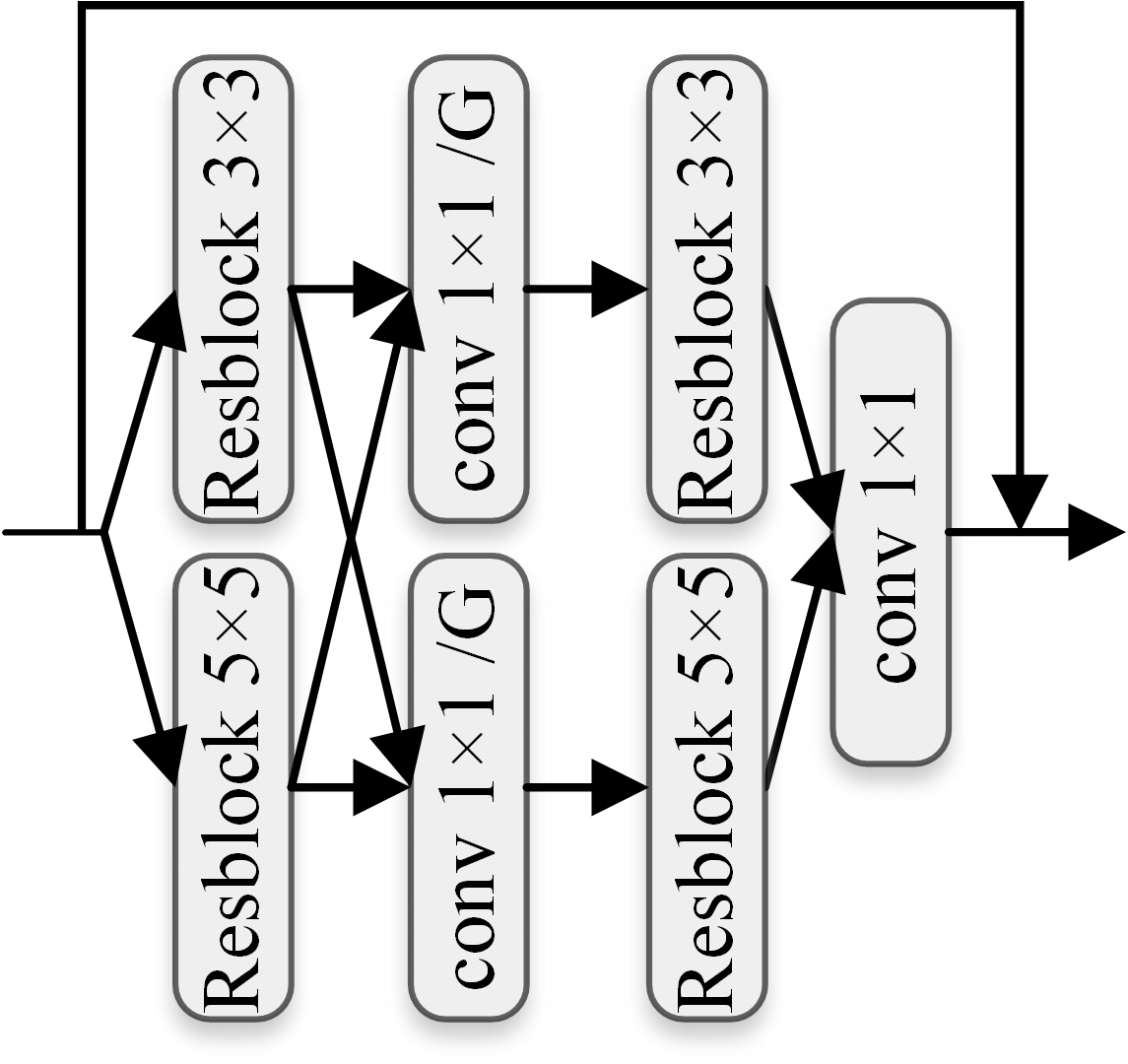}
  \subcaption{IMSRB}
\end{subfigure}
\begin{subfigure}{0.45\textwidth}
  \centering
  \includegraphics[width=0.8\linewidth]{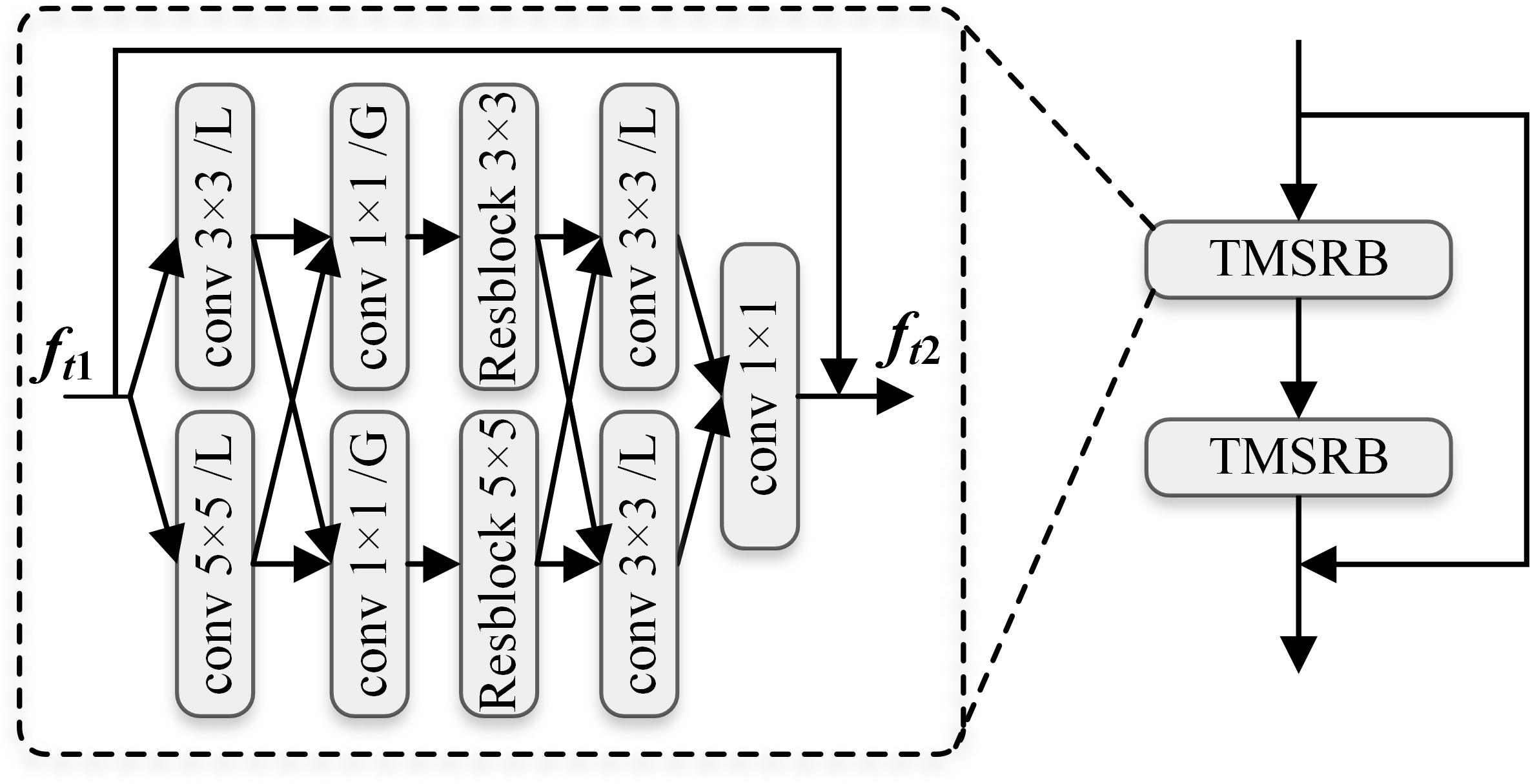}
  \subcaption{CTMSRB}
\end{subfigure}
  \caption{(a) The multi-scale residual block (MSRB) in\cite{li2018multi}. (b) The improved MSRB in\cite{fu2023asymmetric}. (c) The proposed cascaded two-stage multi-scale residual blocks (CTMSRB). $R$ denotes ReLU, $L$ denotes Leaky ReLU, and $G$ denotes GDN.}
\label{fig:CTMSRB}
\vspace{-5pt}
\end{figure}

SC images, with large areas of uniform pixels like webpage backgrounds and sharp edges like text, can benefit from multi-scale networks for diverse scale feature extraction, as depicted in Fig. \ref{fig:CTMSRB}. While Fu {\em et al.}\cite{fu2023asymmetric} improved MSRB using residual block based on \cite{li2018multi}, both approaches still facilitate feature interactions at different scales in a single stage. We introduce a two-stage MSRB (TMSRB) to improve the interaction among multi-scale features. In each TMSRB, features are extracted via two-branch convolutions with different kernel sizes and concatenated for the first stage of feature interaction. Subsequently, the outputs are refined via a residual block before the second interaction phase. The final output is obtained by applying a 1 $\times$ 1 convolution to the concatenated features with a shortcut connection from the original input. Moreover, He {\em et al.}\cite{he2022elic} demonstrated that cascaded residual blocks can replace generalized divisive normalization (GDN) to further enhance nonlinearity and RD performance. Consequently, we construct a cascaded two-stage multi-scale residual block (CTMSRB) to enhance nonlinearity, integrating two TMSRB cascades via shortcut connections, following the output from each ToRB.
\subsection{Window-based attention module}
Leveraging the effectiveness of attention mechanisms in computer vision, we employ the window-based attention block (WAM) \cite{zou2022devil} for image compression, particularly focusing on high-contrast regions in screen content (SC) images. This approach efficiently allocates more bits to complex high-contrast areas while conserving bits in simpler low-contrast regions, with its efficiency validated in Section 4.4.
\subsection{Loss function}
The loss function must effectively account for the joint optimization of the bitrate $r$ and distortion $d$:
\begin{equation}
\begin{aligned}
L= & \lambda \cdot d(\bm{x}, \bm{\hat{x}}) + r_{\bm{y^H}}+r_{\bm{y^L}}+r_{\bm{z^H}}+r_{\bm{z^L}} \\
= & \lambda \cdot d(\bm{x}, \bm{\hat{x}}) + \mathbb{E}\left[-\log _2 p_{\bm{\hat{y}^H}}\left(\bm{\hat{y}^H}\right)\right]+\mathbb{E}\left[-\log _2 p_{\bm{\hat{y}^L}}\left(\bm{\hat{y}^L}\right)\right] \\
& +\mathbb{E}\left[-\log _2 p_{\bm{\hat{z}^H}}\left(\bm{\hat{z}^H}\right)\right]+\mathbb{E}\left[-\log _2 p_{\bm{\hat{z}^L}}\left(\bm{\hat{z}^L}\right)\right],
\end{aligned}
\end{equation}
where $\lambda$ is a Lagrange multiplier, $r$ comprises four terms: $r_{\bm{y^H}}$, $r_{\bm{y^L}}$, $r_{\bm{z^H}}$, and $r_{\bm{z^L}}$, representing the entropy of high-frequency and low-frequency information of latent representation and hyper latent representation, respectively.
\section{Experimental Results}
\label{sec:Experiments}
\subsection{Training dataset}
\begin{figure}
\centering
\includegraphics[width=0.9\linewidth]{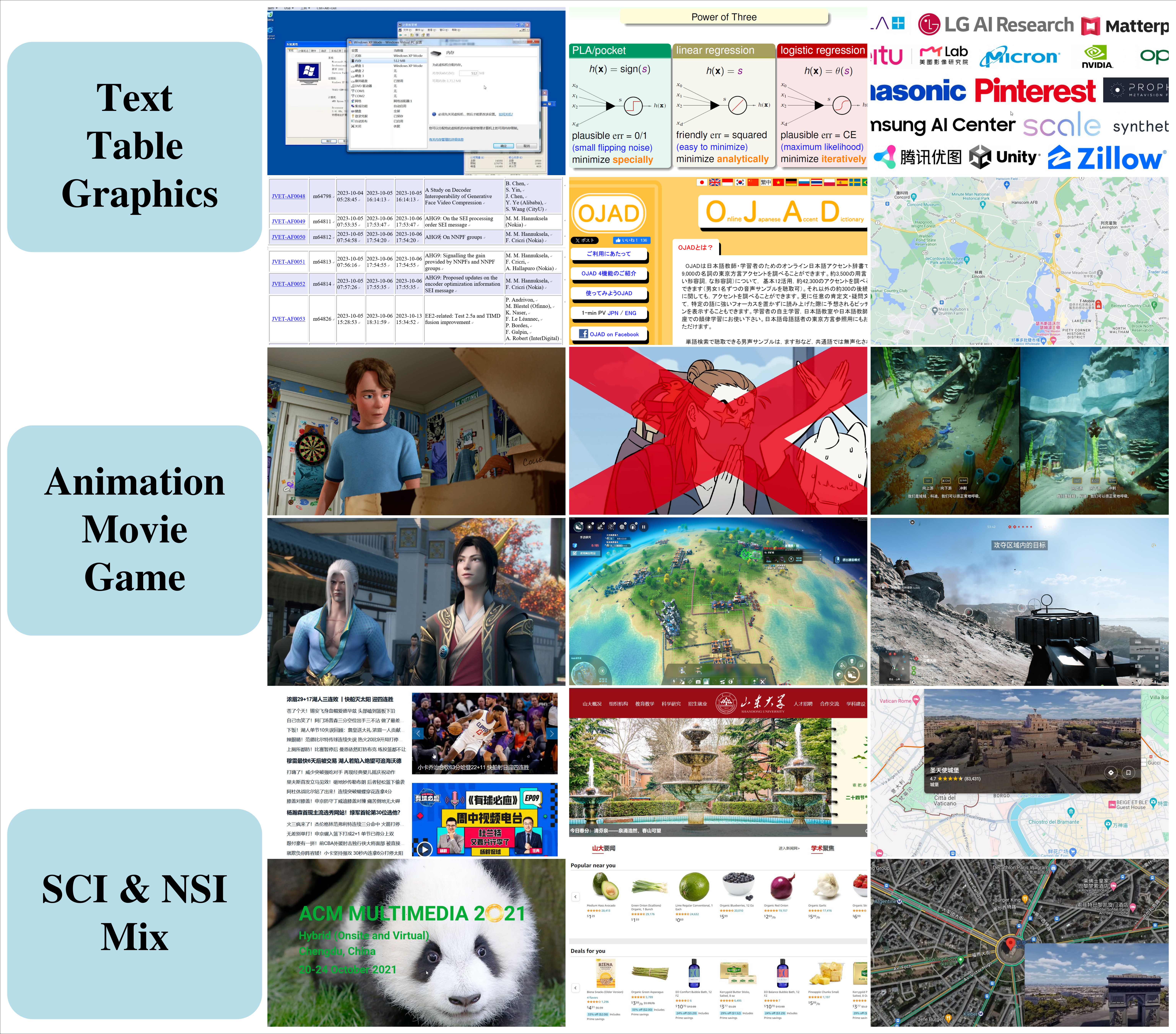}
\caption{Sample images of the proposed SDU-SCICD2K dataset.}
\label{fig:dataset}
\vspace{-5pt}
\end{figure}
Existing SC datasets predominantly emphasize quality assessment~\cite{8468110,9463764} over image compression, containing only a limited number of original SC images inadequate for training compression algorithms. To bridge this gap, we constructed an SC Image Compression Dataset (SDU-SCICD2K), encompassing over 2000 screen content images, as shown in Fig. \ref{fig:dataset}. The dataset was divided into three main categories, each containing approximately 700 images. The first included text, charts, and graphics, the second consisted of animations, movies, and games, and the third combined NS images with SC images. The SDU-SCICD2K is publicly available at https://github.com/SunshineSki/OMR\_Net.git.
\begin{figure}
\begin{minipage}[b]{0.48\linewidth}
  \centering
  \centerline{\epsfig{figure=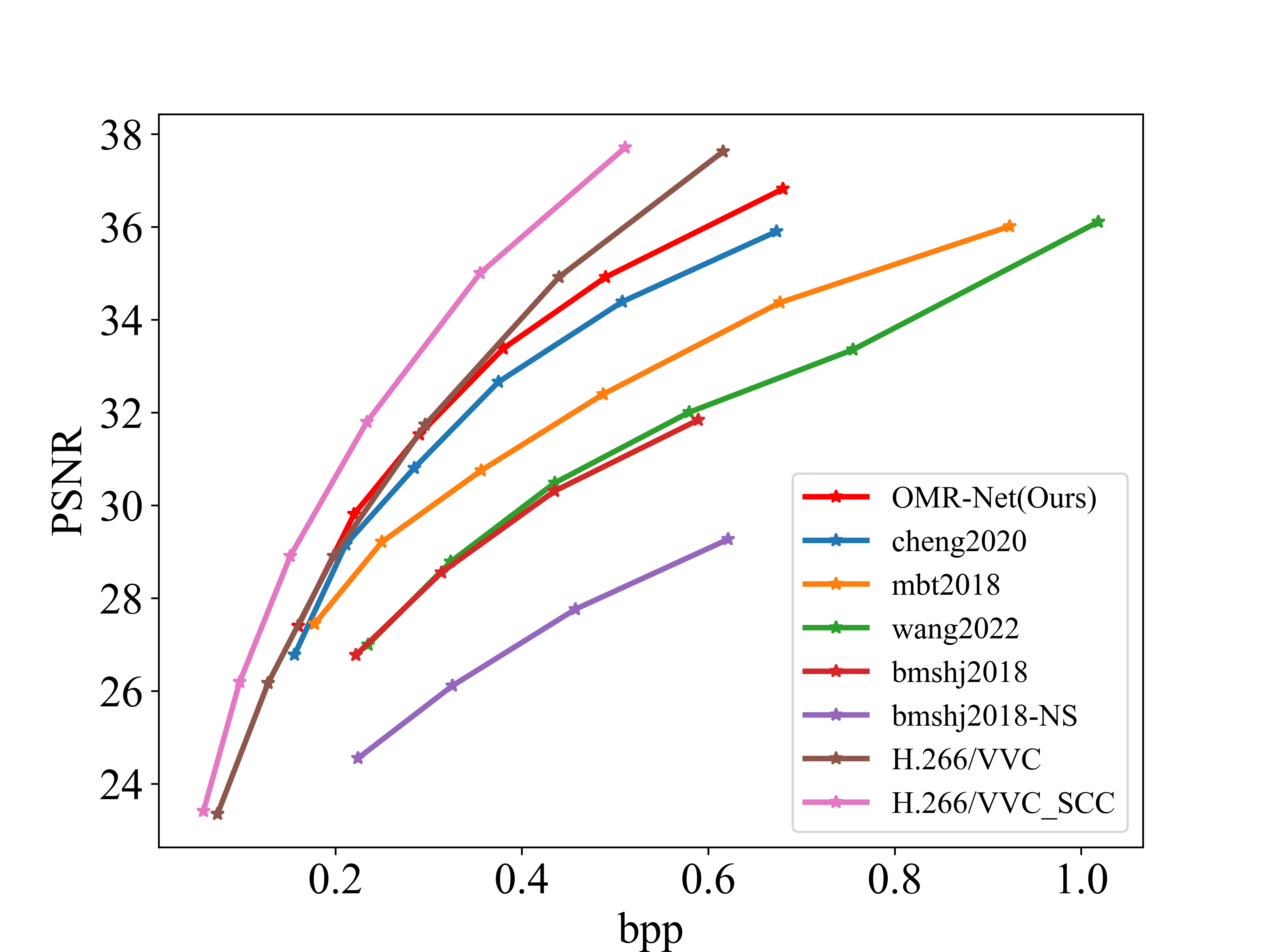,width=4.6cm}}
  \centerline{(a) SCID}\medskip
\end{minipage}
\begin{minipage}[b]{0.48\linewidth}
  \centering
  \centerline{\epsfig{figure=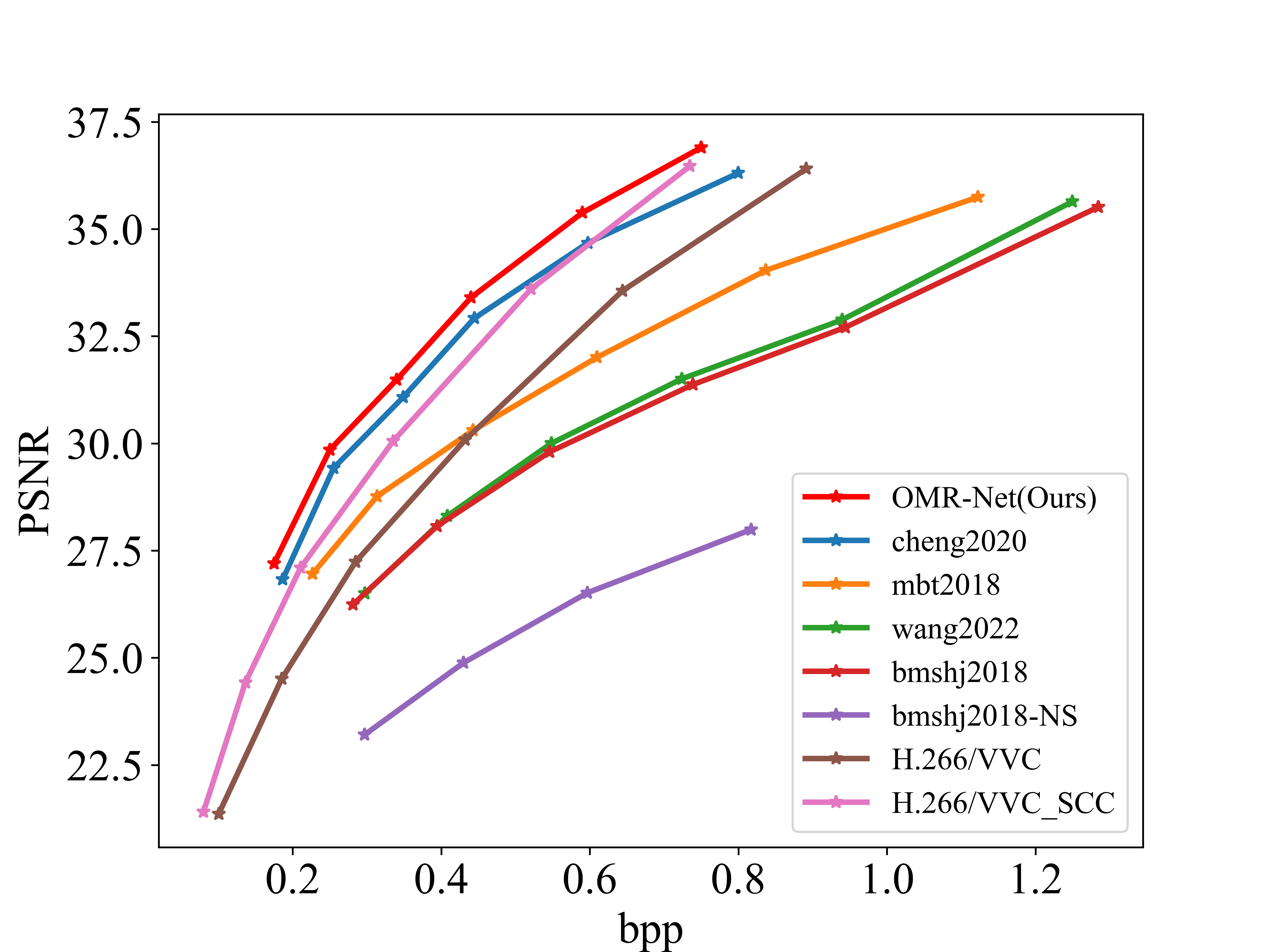,width=4.6cm}}
  \centerline{(b) SIQAD}\medskip
\end{minipage}
\caption{RD Performance evaluation on (a) SCID dataset and (b) the SIQAD dataset.}
\label{fig:rd performance}
\vspace{-5pt}
\end{figure}
\subsection{Experimental setting}
During training, we randomly cropped SDU-SCICD2K images into $256 \times 256$ patches with a batch size of 8. The training process utilized the Adam optimizer for 400 epochs on NVIDIA RTX4090, commencing with an initial learning rate of $10^{-4}$, which was subsequently reduced to $10^{-5}$ at the 300th epoch. We optimized the model with respect to mean squared error (MSE), using $\lambda$ values of \{0.0018, 0.0035, 0.0067, 0.013, 0.025, 0.0483\}, and set $\alpha$ to 0.5. In terms of channel configurations, the encoder and decoder featured 192 convolutional channels for low bit rates and 320 channels for high bit rates. The evaluation was conducted using the well-established SCID~\cite{ni2017esim}, SIQAD~\cite{yang2015perceptual} and CCT~\cite{min2017unified} datasets.
\begin{table}[htbp]
	\centering
	\caption{Average BD-rate(\%) and time complexity. The anchor is `bmshj2018' model.}
	\label{tab:BD-rate}
	\setlength\tabcolsep{3pt} 
	\begin{tabular}{ccccccc}
	\toprule
	\textbf{Method}  & \textbf{SCID} & \textbf{SIQAD} &\textbf{CCT} &\textbf{Kodak} &\textbf{Enc.Time} &\textbf{Dec.Time}\\
	\midrule
	bmshj2018-NS & 75.4 &  116.3 & 145.3& - & 2.51s& 3.42s \\
	wang2022 & -1.8 &  -3.3 & -4.5& -1.2 & 2.69s& 3.58s\\
	mbt2018  &-24.9 &  -25.8 & -29.4& -15.5 & 18.15s& 31.29s\\
	cheng2020  &-43.9 &  -49.7 & -54.3& -23.9 & 14.67s& 34.73s\\
	H.266/VVC  &  -51.9 & -35.3  & -60.2& -25.6 & 82.69s& 0.23s\\
	H.266/VVC\_SCC  & \textbf{-61.8} &  -48.5 & \textbf{-69.3} & -25.1 & 150.09s& 0.24s\\
	\textbf{OMR-Net(Ours)}  &  -50.8 & \textbf{-58.6} & -62.3& \textbf{-28.3}& 17.69s& 36.25s\\
	\bottomrule
	\end{tabular}
\vspace{-8pt}
\end{table}
\subsection{RD performance}
We compared OMR-Net (Ours) with H.266/VVC (VTM19.0), H.266/VVC-SCC~\cite{bross2021overview} (currently the best codec for SC images) and the existing state-of-the-art LIC methods, including cheng2020~\cite{cheng2020learned}, mbt2018~\cite{minnen2018joint}, bmshj2018~\cite{balle2018variational}, and wang2022~\cite{wang2022transform}, implemented within the compressAI~\cite{begaint2020compressai}. We reproduced the experiments of~\cite{wang2022transform} independently. Fig. \ref{fig:rd performance} shows that our method outperforms other LIC methods, exceeding H.266/VVC and H.266/VVC-SCC on SIQAD but lagging behind on SCID. This indicates distribution variations of datasets affects LIC methods' performance. Additionally, bmshj2018-NS model, trained on NS data, demonstrates markedly inferior performance compared to bmshj2018, underscoring the need for SC-specific coding techniques due to data distribution differences. As shown in Table \ref{tab:BD-rate}, the first three columns represent models tested on SCID, SIQAD and CCT, respectively. Furthermore, the results in the fourth column show the performance of models trained on ImageNet and tested on Kodak~\cite{franzen1999kodak}. The last two columns show the encoding and decoding time on SCID. Although improvements still persist in NS images, they are less notable compared to SC images. Using bmshj2018 as an anchor, the proposed method achieves a 28.3\% Bjøntegaard-delta rate (BD-rate) reduction on Kodak, and a significant 50.8\%, 58.6\% and 62.3\% BD-rate reduction on SCID, SIQAD and CCT, respectively. The performance of the proposed method gains notably surpass the other LIC methods, demonstrating its superior suitability for SC image compression.
\begin{figure}
\begin{minipage}[b]{0.48\linewidth}
  \centering
  \centerline{\epsfig{figure=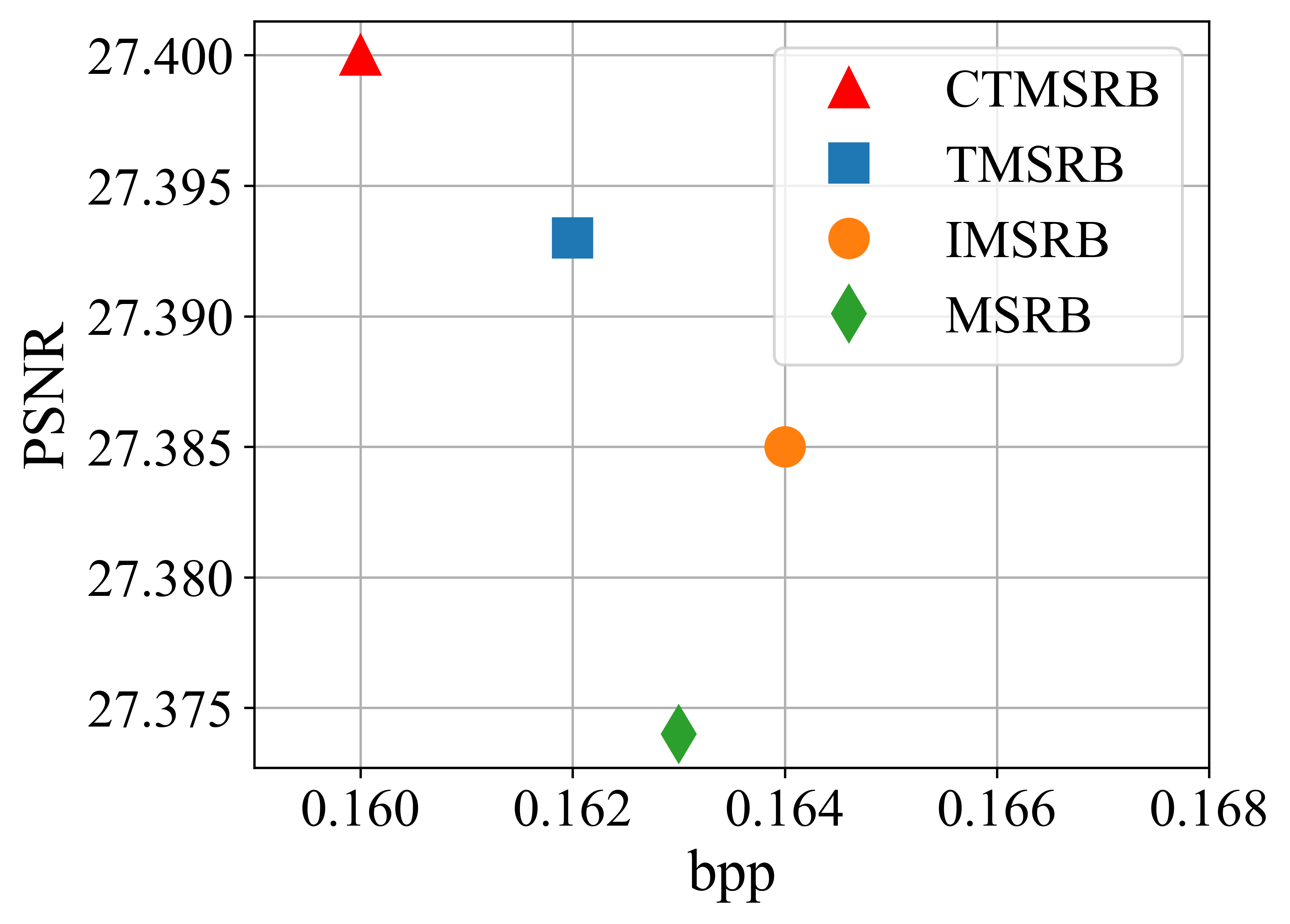,width=4.25cm}}
  \centerline{(a) $\lambda =0.0018$}\medskip
\end{minipage}
\begin{minipage}[b]{0.48\linewidth}
  \centering
  \centerline{\epsfig{figure=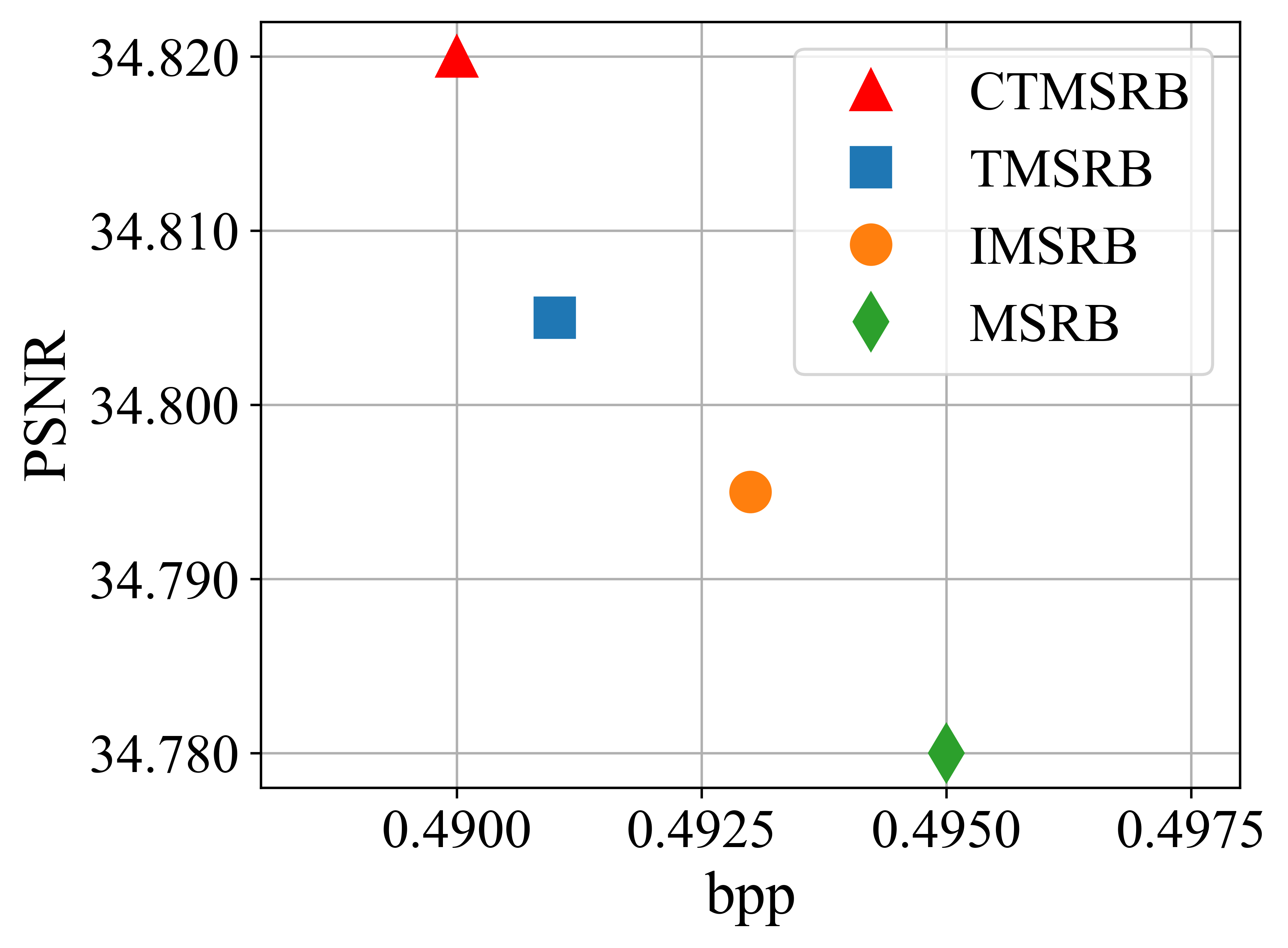,width=4.25cm}}
  \centerline{(b)  $\lambda =0.025$}\medskip
\end{minipage}
\caption{Ablation study on SCID dataset.}
\label{fig:ablation-CTMSRB}
\vspace{-8pt}
\end{figure}
\subsection{Ablation study}
To validate the efficiency of the proposed CTMSRB and WAM, we conducted ablation studies on the SCID dataset, as shown in Table \ref{tab:ablation}. `Basic' indicates the use of the improved ToRB structure combined with GDN, `Basic+CTMSRB' refers to the improved ToRB structure with CTMSRB, and `Basic+CTMSRB+WAM' signifies the proposed method. Compared to the `Basic', the proposed approach achieves a 0.4-0.5 dB increase in PSNR at similar bitrates. The results clearly demonstrate that CTMSRB and AWM significantly enhance the RD performance of our method. Fig. \ref{fig:ablation-CTMSRB} highlights the superiority of the proposed CTMSRB over other MSRB methods, achieving better reconstruction quality at lower bitrates.
\begin{table}
	\caption{Ablation studies. The best performance is highlighted in bold font.}
	\label{tab:ablation}
	\centering
	\begin{tabular}{lccc}
	\toprule
	\textbf{Method} & \textbf{$\lambda$} & \textbf{PSNR$\uparrow$} & \textbf{bpp$\downarrow$} \\
	\midrule
	Basic & 0.0035 & 29.58 & 0.222 \\
	Basic+CTMSRB & 0.0035 & 29.81 & 0.223\\
	\textbf{Basic+CTMSRB+WAM}  & 0.0035 & \textbf{29.99} & \textbf{0.220} \\
	\midrule
	Basic & 0.013 & 32.84 & 0.382\\
	Basic+CTMSRB & 0.013 & 33.18 & 0.384\\
	\textbf{Basic+CTMSRB+WAM} & 0.013 & \textbf{33.39} & \textbf{0.381} \\
	\bottomrule
	\end{tabular}
\vspace{-8pt}
\end{table}
\section{Conclusion}
\label{sec:Conclusion}
In this paper, we introduce a LIC algorithm for SC. We propose an improved ToRB to specifically extract high and low-frequency features of SC. The proposed CTMSRB addresses uniform pixels regions like webpage backgrounds and enhances nonlinear representations. The WAM is employed to direct the network's focus on features in high-contrast regions, essential for sharp edges like text information. Additionally, a specialized image dataset for SC compression is constructed. Experimental results validate the efficacy of the proposed approach, demonstrating superiority over other LIC methods across diverse datasets.

\bibliographystyle{IEEEbib}
\bibliography{strings,refs}
\end{document}